# Compression Algorithm based on Irregular Sequence

Rui Zhu
Second High School Attached to
Beijing Normal University, China
18 Guan'ao Yuan, Baoshengli,
Longgang Road, Qinghe, Haidian
District, Beijing, China
rain_bilibili@outlook.com

## ABSTRACT
The paper introduces a new lossless, highly robust compression algorithm that similar with LZW algorithm, yet the algorithm discards dictionary processing and uses irregular sequences with massive, random information instead. Then the paper found the ineffectiveness of the algorithm due to limited computing ability of hardware and made a few improvements to the algorithm. The algorithm is recommended to be applied in interplanetary communications between a high-compute-ability device and a low-compute-ability receiving device, whose signal would be easily interfered by cosmos rays.

## 1. INTRODUCTION
The algorithm is inspired by the LZW algorithm, a dictionary-based compression algorithm [1-2]. According to Zeng and Rao's research and Zheng's research in 2011, compression algorithms that based on dictionaries would lead to lower compression ratios compared with those that based on arithmetic and statistic methods [3-4]. It is also concluded that the longer the buffer size of dictionary is, the lower the compression ratio will be.

Since the effectiveness of LZW algorithm greatly depends on the size of dictionary, the report is going to introduce a new compression algorithm that uses pre-processed, irregular dictionaries that could have infinitive length, which means they would theoretically contain all data needed, and therefore dictionary processing is no longer required in the algorithm [5]. Root squares of the first 100 prime numbers are used as dictionaries of the algorithm since they are irrational [6].

## 2. CAIS ALGORITHM
### 2.1 Principles

| Define CAIS(target){ |
|---|
| If target.match()==True{ |
|    Return target.baseLocation(); |
| } |
| Else{ |
|    String[] txt= new String[]; |
|    ArrayList buf = new ArrayList<String>; |
|    If target.length()==1{ |
|      Return target; |
|    } |
|    Else{ |
|      txt = target.avgsplit(2); |
|      for i in txt{ |
|         buf.add(CAIS(i)); |
|      } |
|    } |
| } |
| Return buf; |
| } |

**Figure 1. General pseudo-code of CAIS algorithm**

As shown in figure 1, target data would firstly be binarized, and then compared with all base sequences (dictionaries). If match, range of the matched sequence in the certain base sequence and the id of base sequence would be recorded as compressed data; if not, the target sequence would be continuously split into several shorter subsequences until every sequence is matched. If a subsequence (character) could no longer be split and still could not be matched, the subsequence (character) itself would be recorded.

**Target Sequence**
1011101110111010000 10011

**Base Sequence**
11 | 1011101110111010000 10011 | 11
23                                               46

**Figure 2. Algorithm principle**

For example, as shown in figure 2, the target sequence matches the base sequence from number 23 to number 46, thus the target sequence could be referred as "23, 46" for the base sequence. Therefore, the more the target sequence corresponds to the base sequence, the lower the compression ratio would be.

### 2.2 Expectation
Since the irregular sequences, which are used as dictionaries in the algorithm, can be calculated as long as possible, every kinds of target sequence could be regarded as a subsequence of a base sequence, and therefore any data could be expressed as a sequence consists of 3 elements: the number of the base sequence, start index of the target sequence on base sequence and the end index. It would be an incredibly effective data compression algorithm.

## 2.3 Results

Despite the algorithm would theoretically compress data in a low compression ratio, the test results contradict with the expectation. To be specific, the results are the reverse of the theory predicts.

**Table 1. Results of 10 short texts**

| Text# / Category | 1 | 2 | 3 | 4 | 5 | 6 | 7 | 8 | 9 | 10 |
|---|---|---|---|---|---|---|---|---|---|---|
| Matched # | 9 | 0 | 9 | 7 | 7 | 9 | 10 | 8 | 6 | 12 |
| Original length | 15 | 4 | 14 | 11 | 9 | 14 | 16 | 12 | 8 | 19 |
| Result length | 68 | 4 | 61 | 42 | 34 | 51 | 72 | 24 | 12 | 83 |

As shown in table 1, the algorithm could not find any match for several target sequences, and it seems that the number of matched characters would increase as the length of target sequence increases. Moreover, the length of results is always longer than that of the input sequence. It should be the irrationality of the base sequences that leads to such poor matching result and high compression ratio.

## 2.4 Improvements

### 2.4.1 Filter Strategy

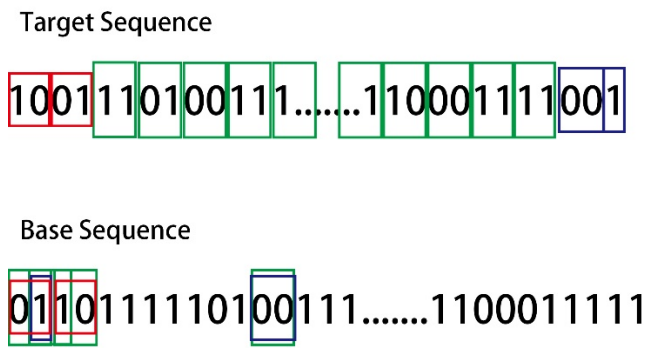

**Figure 3. Example of regular CAIS result**

Figure 3 shows a common example of a target sequence after processed by normal CAIS algorithm. The matched fragments are always too short and represent the first few substrings of base sequence only.

For increasing the efficiency of matching, filter strategy is created and applied. In phase one of filter strategy, each pair of adjacent matched fragments of the target sequence will be combined and then compared with the base sequence, and numeral strings represent the matched pair would cover the original values that indicate it. In each round except the first, fragments that cannot be matched when combined with adjacent ones in previous rounds would not be included. This process would be repeated until no new combinations are matched or no new combinations could be made.

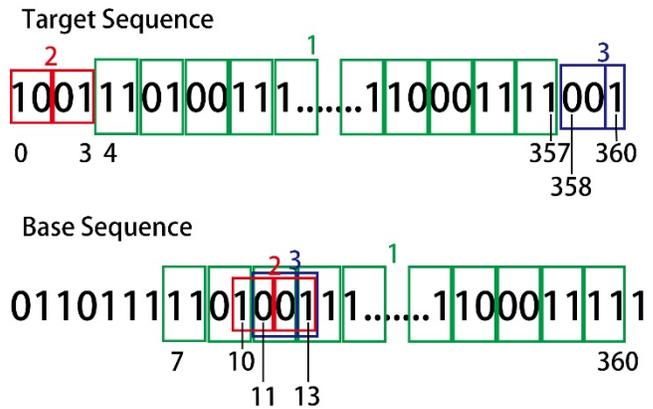

**Figure 4. Result of the example after phase one of filter strategy**

As shown in figure 4, after phase one of filter strategy, fragments of the target sequence have been combined and matched to a longer and consecutive subsequence of the base sequence, which is a big progress on improving the efficiency of the algorithm.

To make another progress on algorithm effectiveness, another phase of data processing is applied.

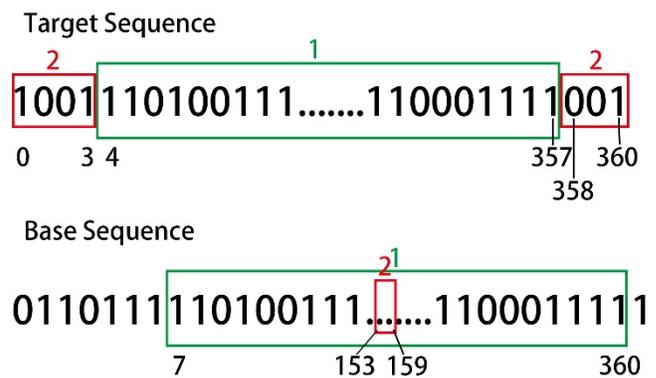

**Figure 5. Result of filter strategy phase two**

In phase two, the algorithm would temporarily extract the longest matched data subsequence, and then repeat part of phase 1 for the rest, for most digits would have already matched and combined as chunks, and the new data sequence would be processed by filter strategy phase two again, until there is only one matched subsequence left or when there are several subsequences in the same length.

Matched data segments whose position sequences are longer than the segments themselves would remain their original values in the result.

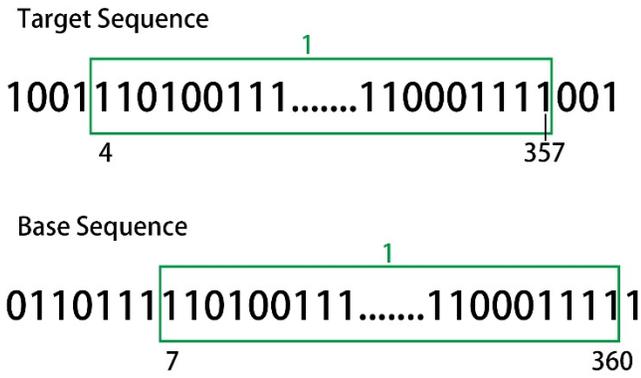

**Figure 6. Result of filter strategy**

As shown in figure 6, this strategy would make sure that the result would at least not longer than the original text.

**Table 2. Effectiveness comparison between using filter strategy and not**

| Algorithm  Category | Normal CAIS | CAIS using filter strategy |
|---|---|---|
| Matched# | 659 | |
| Original length | 910 | |
| Result length | 1587 | 896 |

For the first time, the result sequence of the algorithm is smaller than its input sequence, which proves the strategy's effectiveness. The reason that such great difference occurred might be that the longest substring has been split into several substrings and thus cost more spaces to indicate the information. Moreover, many matched sequences that are too short to have a compression rate lower than 100% are deleted and recorded as the original subsequence instead.

### 2.4.2 LCS Strategy

For further improvement, the longest common substring of target sequence and the base sequence would be exerted from target sequence each time, until every character of the target sequence have been indicated. The method of finding the longest common string (LCS) is adopted from Babenko and Starikovskaya's research in 2011 [7].

**Table 3. Effectiveness comparison of filter strategy and LCS strategy**

| Algorithm  Category | CAIS with LCS strategy | CAIS with filter strategy |
|---|---|---|
| Matched# | 659 | |
| Original length | 910 | |
| Result length | 1223 | 896 |

| Time (around) | 2 hours and 26 minutes | 47 minutes |
|---|---|---|

As indicated in table 3, CAIS with LCS strategy is cost more time and output longer result than CAIS with filter strategy since CAIS with LCS does not remove information of useless matched sequences and has a higher time complexity. In conclusion, CAIS with filter strategy would be the better solution.

### 2.4.3 Extending Base Sequence

**Table 4. Comparison of the percentages of match of the same target sequence in difference length of base sequences**

| Number of decimals | 50000 | 100000 | 500000 |
|---|---|---|---|
| Match percentage | 72.42% | 80.77% | 99.12% |

As presented in Table 4, the percent of match would grow as the base sequences enlarges. For single-character match percentage is associated with compression ratio, the compression ratio of the algorithm would be improved as well.

In conclusion, using CAIS with filter strategy and base sequences as long as possible would significantly improve the algorithm's effectiveness.

## 3. EVALUATION AND DISCUSSION

**Table 5. Comparison on compression ratios of CAIS and that of other common algorithms on the same datasets**

| | LZW | LZW18 [8] | CAIS | CAIS using filter | CAIS using LCS |
|---|---|---|---|---|---|
| numeral | 180% | 180% | 253% | 100% | 89% |
| English text | 202% | 202% | 502% | 100% | 673% |
| UTF-8 text | 145% | 144% | 622% | 100% | 738% |

(continue)

| | CAIS using filter and bases in 500000 decimals | Huffman Tree [9] |
|---|---|---|
| numeral | 32% | 61% |
| English text | 100% | 98% |
| UTF-8 text | 100% | 97% |

Note: all datasets <150 bytes.

As indicated in table 5, the improved CAIS is more effective on handling numeral data, yet not that effective on English (ascii) data and UTF-8 data. Using a UTF-8 base sequence full of regular used words and sentences might be increase effectiveness of the algorithm on UTF-8 text compression.

**Table 6. Comparison on compression ratios and match rate of CAIS using numeral and Dureader[1] as base sequence**

Link: https://ai.baidu.com/broad/subordinate?dataset=dureader

---

[1] Dureader is a largescale Chinese Machine Reading Comprehension dataset collected by Baidu, Inc.

| (compression rate/match rate) | CAIS using filter strategy | CAIS using filter strategy and Dureader as base sequence |
|---|---|---|
| Chinese text 1 | 100%/25% | 81%/100% |
| Chinese text 2 | 100%/8% | 100%/100% |
| Chinese text 3 | 100%/23% | 100%/96% |

As indicated in table 6, the solution that used Dureader as base sequence has made a significant increase on match rate and a slight improvement on compression rate. Therefore, using a specified data sequence as an alternative of base sequence would increase efficiency in specified environment.

To evaluate time spent on compression, a set of five 100-byte English text are compressed by five compression algorithms: CAIS, LZW18, CAIS using filter strategy, CAIS using LCS and Huffman tree. Average time cost of the five trials are then compared as table 7.

**Table 7. Comparison on time for encode and decoding of compression algorithms**

|  | CAIS | LZW18 [8] | CAIS using filter | CAIS using LCS | Huffman Tree [9] |
|---|---|---|---|---|---|
| Average encode time | 2 min 54 sec | 1.62 sec | 14 min 37 sec | 1 hr 49 min 17 sec[2] | 1.18 sec |
| Average decode time | 1.43 sec | 0.48 sec | 1.98 sec | 1.86 sec[3] | 1.01 sec |

Table 7 has proved the previous assumption that CAIS would cost much less time on decoding than time spent on encoding process. However, CAIS still needs more time to decode compared with other compression algorithms, which may lead to disadvantages on decoding massive data. Both encoding and decoding can be improved by applying more efficient computing strategies, such as multithread and Single Instruction Multiple Data.

## 4. CONCLUSION
Though still has a high compression ratio, it could be concluded that CAIS could process much shorter results if longer base sequences are calculated and applied as dictionaries in the algorithm since the irrationality of square root value would remain constant. Furthermore, since the time and space complexity of the algorithm would exponentially increase as the length of base sequence increases, CAIS Algorithm would maximize its ability when it is used to send data in highly interfering circumstances that both side of communication are devices with high computing capacity, such as communications between NASA Ground Control on Earth and Curious on Mars.

The algorithm could be improved by improving matching strategies, computing strategies and extending base sequences. If possible, constitution of base sequences could be modified to adapt specific applicative environments. For example, using massive texts from Wikipedia as base sequences of CAIS would greatly improve its effectiveness in English communications.

---

[2] Only measured for twice, for it cost too long on encoding. Hence the data might not be accurate.

[3] Measured by the only two successful results.